\documentclass[aps,pre,preprint,groupedaddress]{revtex4}

\usepackage{graphicx}
\usepackage{latexsym}
\usepackage{amsmath,amssymb}

\begin{document}

%\preprint{}

\title{
Transfer-matrix approach to the three-dimensional bond percolation:
An application of Novotny's formalism
}

\author{Yoshihiro Nishiyama}
%\thanks{}
\affiliation{Department of Physics, Faculty of Science,
Okayama University, Okayama 700-8530, Japan.}

\date{\today}

\begin{abstract}
A transfer-matrix simulation scheme for the three-dimensional 
($d=3$) bond percolation is presented.
Our scheme is based on Novotny's transfer-matrix formalism,
which enables us to consider arbitrary (integral) number of sites
$N$ constituting a unit of the transfer-matrix slice even for $d=3$.
Such an arbitrariness allows us to perform systematic finite-size-scaling analysis
of the criticality at the percolation threshold.
Diagonalizing the transfer matrix for $N =4,5,\dots,10$,
we obtain an estimate for the correlation-length critical exponent 
$\nu = 0.81(5)$.
\end{abstract}

%\keywords{}

\maketitle

\section{\label{section1}Introduction}

The transfer-matrix method has an advantage over the Monte Carlo method
in that it provides information
free from the statistical (sampling) error and the problem of slow relaxation
to thermal equilibrium.
On one hand, 
the tractable system size
is severely limited, because the transfer-matrix size increases exponentially
as the system size enlarges.
Such a difficulty could be even more serious for ``geometrical'' problems 
such as the percolation,
for which the configuration space is much larger than that of the
Ising model, for example.

A transfer-matrix approach to the percolation in two dimensions ($d=2$) 
was initiated by Derrida and Vannimenus \cite{Derrida80}.
They treated the system sizes (transfer-matrix strip widths) up to $N=5$.
Performing an extensive phenomenological renormalization group 
(finite-size scaling) analysis \cite{Nightingale76},
they estimated the correlation-length critical exponent 
as
$\nu=1.2 \sim 1.4$; the variance is due to the choice of the
boundary conditions.
Their result is
quite consistent with the exact value $\nu=4/3$,
indicating that the transfer-matrix approach to percolation would be promising.
Because the transfer-matrix data are free from the statistical error,
the data allow us to take its numerical derivative,
which provides valuable information as to the subsequent
finite-size-scaling analysis.

It turned out, however, that
its naive extension to the $d=3$ case is rather problematic;
we refer to Sec. 4.4 of Ref. \cite{Stauffer94} for an overview.
Actually, for $d=3$, as the system size (linear dimension)
$L$ enlarges, the number of constituent sites $N(=L^2)$ 
of the transfer-matrix unit soon exceeds
the limit of the available computer resources.

The aim of this paper is to develop an improved 
version of the transfer-matrix formalism for the $d=3$ bond percolation.
For that purpose,
we utilize Novotny's idea,
which has been applied successfully to various Ising models in $d \le 7$
\cite{Novotny90,Novotny92,Novotny93,Novotny91,
Nishiyama04,Nishiyama05}.
His formalism stems on a very formal expression for 
the transfer-matrix elements.
It enables us to consider arbitrary (integral) number of 
constituent sites ${}^\forall N$
even for $d \ge 3$.
Owing to this arbitrariness, we are able to treat a variety of system sizes,
and
perform systematic finite-size scaling analysis of the numerical data.
In this paper, we demonstrate
that his idea is applicable to the 
$d=3$ bond percolation.

The rest of this paper is organized as follows.
In Sec. \ref{section2}, we formulate the transfer-matrix scheme
for the $d=3$ bond percolation.
We place
an emphasis how we adapted Novotny's idea to 
the bond-percolation problem.
The simulation results are shown in Sec. \ref{section3}.
Taking an advantage that we can treat various system sizes,
we manage the phenomenological renormalization group \cite{Nightingale76}
(finite-size scaling) analysis.
Thereby, we obtain an estimate for the correlation-length
critical exponent $\nu=0.81(5)$.
In Sec. \ref{section4}, we present the summary and discussions.

\section{\label{section2}
Construction of the transfer matrix for the 
three-dimensional bond percolation 
}

In this section, we set up the transfer-matrix formalism for the $d=3$ bond percolation.
As mentioned in Introduction, we adopt Novotny's idea \cite{Novotny90}.
So far, his idea has been applied to various types of the Ising models in $d \le 7$.
Here, we show that his idea is also applicable to the bond percolation.
At the end of this section, we argue a conceptual difference from 
the original Novotny method.

\subsection{Configuration space}

Above all, we need to set up the configuration space
so as to represent the transfer-matrix elements explicitly.
The bases of the configuration space should specify all possible connectivities among the $N$ sites
which constitute a unit of the transfer matrix, namely,
a cross section of the transfer-matrix bar; see Fig. \ref{figure1}.
In the figure, we also presented a drawing of an example of connectivity among 
the $N=4$ sites.

As shown in the figure, an integer index $\alpha=1,2,\dots,N$ specifies
the position of the constituent
sites of a transfer-matrix unit.
In order words, the transfer-matrix unit is of one-dimensional structure rather
than two-dimensional one.
Such a feature might be confusing, compared with the drawing in Fig. \ref{figure1} (a),
where the
transfer-matrix unit is drawn as a rectangular shape with the edges $\sqrt{N}$.
Actually, the dimensionality is lifted to $d=2$ afterward by introducing 
the $\sqrt{N}$th-neighbor long-range interactions among the $N$ sites.
We will explain this scheme explicitly in the next subsection.
Here,
for the time being, we consider that these $N$ sites are arranged into
a one-dimensional structure.

In order to specify the connectivity among the $N$ sites,
we accepted the following matrix-based representation;
\begin{equation}
 [a_i]_{\alpha \beta} = 
\left\{
\begin{array}{ll}
1 & for \ a \  pair \  of \  connected \  sites \  (\alpha,\beta)   \\
0 & otherwise
\end{array}
\right.         .
\end{equation}    
The index $i$ runs over all possible connectivities among the $N$ sites.
For example, the connectivity of Fig. \ref{figure1} (b) is 
represented by,
\begin{equation}
\label{matrix_representation}
a=
\left(
\begin{array}{cccc}
1 & 1 & 0 & 1 \\
1 & 1 & 0 & 1 \\
0 & 0 & 1 & 0 \\
1 & 1 & 0 & 1
\end{array}
\right)   .
\end{equation}

Let us mention a few remarks:
The authors in Ref. \cite{Derrida80}, accepted more elaborated representation scheme.
Namely, they specified whether a cluster is connected with the ``origin'' 
or that a cluster is an isolated one.
Here, the ``origin'' stands for an edge of the transfer-matrix bar,
An advantage of their extended representation is that one only needs to evaluate
the largest eigenvalue of the transfer matrix to obtain the correlation length.
Here, however,
we did not accept their representation scheme.
Correspondingly, we calculated the sub-dominant eigenvalue together with
the dominant one in order to calculate the correlation length.
This task is not so computationally demanding, and 
it renders significant simplification of the formalism
mentioned below.

\subsection{Explicit formula of the transfer-matrix elements 
for the $d=3$ bond percolation}

In this subsection, we present the explicit formula for the
transfer-matrix elements.
We consider an anisotropic bond percolation on the cubic lattice.
Namely,
we set the percolation probabilities $p_\parallel$, $p_{\perp 1}$ 
and $p_{\perp 2}$ independently
for the respective bond directions along the $d=3$ Cartesian axes.
Correspondingly,
we factorize the transfer matrix into the following three components;
\begin{equation}
T=T_\perp(v,p_{\perp 2})  T_\perp(1,p_{\perp 1})  T_\parallel (p_\parallel)  ,
\end{equation}
with,
\begin{equation}
v=\sqrt{N}         .
\end{equation} 
(We followed the idea of Derrida and Vannimenus,
who decomposed the transfer-matrix for the $d=2$ percolation into two factors
\cite{Derrida80}.)
% Here, the symbol $\odot$ denotes the Hadamard
% (element by element) matrix multiplication.
The components $T_\parallel(p_\parallel)$,
$T_\perp(1,p_{\perp 1})$ and $T_\perp(v,p_{\perp 2})$ denote 
the transition probabilities due to 
the longitudinal bond, intra-cluster nearest-neighbor bond,
and the intra-cluster $v$th-neighbor bond, respectively.
(Here, the ``cluster'' stands for the $N$ sites constituting a unit of the transfer-matrix slice,
and the ``longitudinal'' direction is parallel to the transfer-matrix bar; see Fig. \ref{figure1}.) 

In other words, the product of two components 
$T_\parallel(p_\parallel) T_\perp(1,p_{\perp 1})$, namely, with $T_\perp(v,p_{\perp 2})$
ignored,
should yield the transfer matrix
for the $d=2$ bond percolation.
The remaining factor $T_\perp(v,p_{\perp 2})$ lifts
the dimensionality to $d=3$.
%(Note that the parameter $v(=\sqrt{N})$ is not necessarily an integral number.)

We present the explicit formulas for each component.
First, for simplicity, we consider the longitudinal part
$T_\parallel$.
(This is essentially the same as the horizontal factor $M_H$ appearing
in the formalism \cite{Derrida80} for the $d=2$ percolation.)
Our formula for the elements of $T_\parallel$ is given by,
\begin{equation}
\label{transfer_matrix_parallel}
[ T_\parallel (p_\parallel) ]_{ij} =\sum_{ \{J_\alpha \} } p( \{J_\alpha\} , p_\parallel) 
        \left(  a_i , m( \{ J_\alpha \} ) \otimes a_j  \right)
  .
\end{equation}    
Here, the summation $\sum_{ \{ J_\alpha \} }$ runs over all possible random-bond
configurations $\{ J_\alpha \}$ with either $J_\alpha=0$ (unoccupied bond) 
or $1$ (occupied bond) for
$\alpha=1,2,\dots,N$.
The probability $p( \{ J_\alpha \} ,p_\parallel)$ is given by,
\begin{equation}
P( \{ J_\alpha \} ,p_\parallel)= p_\parallel^{N_{J=1}} (1-p_\parallel)^{N-N_{J=1}} 
,
\end{equation}   
with the number of occupied bonds $N_{J=1}$.
The ``random bond'' matrix $m( \{ J_\alpha \} )$ is a diagonal 
$N \times N$ matrix with the 
diagonal elements 
$ [ m(\{ J_\alpha \}) ]_{\beta \beta }=J_\beta$.
The operation $\otimes$ denotes the matrix product,
\begin{equation}
[a \otimes b]_{\alpha \beta}=\sum_{\gamma=1}^N  {}'a_{\alpha \gamma} \wedge b_{\gamma \beta}
,
\end{equation}
with the logical product $\wedge$ in the Boolean algebra.
(The summation $\sum'$ gives $1$, unless all the summands are zero.)
The product $(a_i,a_j)$ accounts for the orthogonality of the matrices; namely,
\begin{equation}
(a_i,a_j)=\delta_{ij}  ,
\end{equation}
with Kronecker's symbol $\delta_{ij}$.

As would be apparent from the above formula (\ref{transfer_matrix_parallel}), 
the transfer-matrix element $[T_\parallel]_{ij}$
stands for the transition probability from the initial configuration
$a_j$ to the final configuration $a_i$ through the 
longitudinal random-bond percolation $\{ J_\alpha \}$.

Second, we turn to considering the transverse component $T_\perp(w,p_\perp)$.
This component accounts for the intra-cluster $w$th-neighbor random-bond 
percolation with
the percolation probability $p_\perp$.
This factor is the most significant part in our formalism.
We propose the following formula for $T_\perp (w,p_\perp)$;
\begin{equation}
[T_\perp(w,p_\perp) ]_{ij} =\sum_{ \{ J_\alpha \} } 
  p(\{J_\alpha \},p_\perp) t_{ij}(w, \{ J_\alpha \} )
   .
\end{equation}   
The transition amplitude $t_{ij}$ is given by, 
\begin{equation}
\label{perp_part}
t_{ij}(w, \{ J_\alpha \} ) 
= \sum_{\beta=1}^N  f_\beta (w)  \left( a_i , m_\beta( \{ J_\alpha  \} )
   \oplus a_j \right)   .
\end{equation}   
Here, the symbol $\oplus$ denotes an operation,
\begin{equation}
[a \oplus b]_{\alpha \beta}=\sum_{\gamma=1}^N {}' a_{\alpha \gamma} \vee b_{\gamma \alpha}  ,
\end{equation} 
with the logical summation $\vee$ in the Boolean algebra.
The ``$\beta$th-neighbor-random-bond matrix" $m_\beta (\{J_\alpha \})$ is given by
the formula,
\begin{equation}
m_\beta( \{ J_\alpha \} ) = m(\{ J_\alpha \}) \otimes s_\beta     , 
\end{equation}
with the shift operator,
\begin{equation}
[s_\beta]_{\gamma \delta} = 
\left\{
\begin{array}{ll}
1 & for \  \gamma-\delta = \beta \  mod \ N   \\
0  & otherwise
\end{array}
\right.              .
\end{equation}
The operation $s_\beta$ shifts the diagonal random-bond operator $m( \{J_\alpha \} )$
to an off-diagonal one,
which now represents the $\beta$th-neighbor random bonds.
That is, the operation
$m_\beta( \{ J_\alpha \})  \oplus a_j$ introduces new
intra-cluster $\beta$th-neighbor-random-bond percolation
in adding to the initial connectivity $a_j$.

In order to implement the $w$th-neighbor intra-cluster percolation with a non-integral
value of $w$, 
we need to average over all sectors $\beta=1,2,\dots,N$ with an
appropriate weight $f_\beta(w)$; see Eq. (\ref{perp_part}).
We propose that the weight should be
given by the $w$th-order power of the shift operator;
\begin{equation}
f_\beta (w)=  [(s_1)^w]_{1 \beta}   .
\end{equation}
As would be apparent from the definition, the operator $(s_1)^w$
generates the translational shift of the distance $w$,
and the factor $f_\beta(w)$ picks up the amplitude of each sector $\beta$.
Hence, the resulting formula, Eq. (\ref{perp_part}), should be the transition
amplitude from $a_j$ to $a_i$ by the $w$th-neighbor random-bond
percolation.
Hence, the product $T_\perp(v,p_{\perp 2}) T_\perp(1,p_{\perp 1})$ with $v=\sqrt{N}$
introduces a two-dimensional intra-cluster percolation among the $N$ sites.
As noted previously, a crucial point is that there is no restriction to the number
of constituent sites $N$.

Lastly, let us argue a conceptual difference from the original Novotny method
\cite{Novotny90} for the Ising ferromagnet.
In the original method, the translation operator $(s_1)^w$
acts on the configuration space, which has huge dimensionality.
On the contrary, in our formalism,
the operator $(s_1)^w$ is a mere $N \times N$ matrix.
Hence, from the viewpoint of the computer programing,
the present formalism for the percolation 
is even simpler than the original method.
(Here, the generation of the list of connectivity $\{ a_i \}$
is the most time-consuming.)

\section{\label{section3}Numerical results}

In the above section, we set up the transfer-matrix formalism for the $d=3$
bond percolation.
In this section, we present the numerical results by means of the exact 
diagonalization of the transfer matrix.
We consider the anisotropic bond percolation.
The anisotropy parameter $R$ governs the mutual ratios of the percolation probabilities;
namely,
\begin{equation}
R^{-1} p_\parallel = R^{-1} p_{\perp 1} = p_{\perp 2} =p .
\end{equation}
We consider the anisotropy ratio $R$ as a freely tunable parameter to
stabilize the finite-size corrections.
Diagonalizing the system sizes $N=4,5,\dots,10$, we analyze the percolation transition
in terms of the phenomenological renormalization group \cite{Nightingale76} method.
Note that the linear dimension of the system $L$ is given by the relation,
\begin{equation}
\label{L_and_N}
L=\sqrt{N} ,
\end{equation}
as shown in Fig. \ref{figure1}.

\subsection{Percolation threshold $p_c$}

In Fig. \ref{figure2}, we plotted the scaled correlation length $\xi/L$ for the
percolation probability $p$ with the fixed anisotropy parameter $R=3.3$.
(Afterward, we explain how we adjusted the anisotropy parameter to $R=3.3$.)
The symbols 
$+$, $\times$, $*$, $\Box$, $\blacksquare$,  $\circ$, and $\bullet$
denote the system sizes $N=4$, $5$, $6$, $7$, $8$, $9$, and $10$, respectively.
The correlation length $\xi$ is calculated by the formula
$\xi= 1/\ln (\lambda_1/\lambda_2)$ with the dominant $\lambda_1$ and
the sub-dominant $\lambda_2$ eigenvalues of the transfer matrix.

The intersection point of the curves in Fig. \ref{figure2} indicates the location of the
critical point (percolation threshold) $p_c$.
(The scaled correlation length $\xi/L$ is invariant 
with respect to the system size $N$ at $p=p_c$.)
Hence, we observe that a percolation transition occurs around $p \approx 0.096$.

In order to determine the critical point $p_c$ more precisely,
in Fig. \ref{figure3}, we plotted the approximate transition point 
$p_c(L_1,L_2)$ for $(2/(L_1+L_2))^2$.
Here, the approximate transition point $p_c(L_1,L_2)$ denotes 
the location of the intersection point of the curves 
for a pair of $( L_1 , L_2 ) $.
The symbols $+$, $\times$, and $*$ show that the differences of the 
system sizes are $N_1-N_2=1$, $2$, and $3$, respectively; note that the relation
$L_{1,2}=\sqrt{N_{1,2}}$ holds.
As indicated in Fig. \ref{figure3}, we survey several values of the anisotropy parameter 
such as
$R=2.8$, $3.3$ and $3.8$.
Thereby,
we notice that the finite-size corrections to $p_c$ are suppressed 
on setting $R=3.3$.
The least-square fit to the data for $R=3.3$ yields the critical point $p_c=0.0958(27)$ in
the thermodynamic limit $L\to\infty$.

Let us argue the role of the anisotropy parameter $R$.
First of all, it is worthwhile that the system size 
along the transfer-matrix (longitudinal)
direction is infinite, whereas the transverse system sizes are
both finite $L \le \sqrt{10}$; see Fig. \ref{figure1}.
In this sense, it is by no means
necessary to consider the isotropic condition $p_{\perp1}=p_{\perp2}=p_\parallel$
specifically.
Hence, we consider that
the ratio $R=p_\parallel/p_{\perp2}$ is a tunable parameter.
Practically, we found that
the finite-size corrections improve for large $R$.
Second, we need to remedy the dimensionality $d=3$ by adjusting the ratio 
of the intra-cluster interactions $R=p_{\perp1}/p_{\perp2}$.
That is,
the ``effective dimension'' \cite{Novotny92,Novotny93} can 
deviate slightly from $d=3$,
at least, for small system sizes.
(This deviation deteriorates the finite-size-scaling analysis.)
Note that basically, the backbone structure of Novotny's transfer matrix
is of $d=2$,
and the dimensionality is lifted
to $d=3$ by introducing the long-range interactions among the
intra-cluster sites.
In other words, it is not quite obvious that the dimensionality $d=3$ is realized precisely,
at least, for small $N$.
Hence, in order to remedy this dimensionality deviation,
we should tune \cite{Novotny92,Nishiyama05}
the intra-cluster-interaction ratio $R=p_{\perp1}/p_{\perp2}$.
(Note that for large $R$, 
the component 
$T_\perp(1,p_{\perp1})$ dominates $T_\perp(v,p_{\perp2})$,
and the dimensionality reduces to $d=2$.
For a certain moderate value of $R$, the dimensionality would approach $d=3$.)
More specifically, we adjusted $R$ so as to improve the finite-size-scaling behaviors
of $p_c(L_1,L_2)$ \cite{Novotny92,Novotny93}
as shown in Fig. \ref{figure3}.
In this respect, there might exist alternative parameterization schemes
other than the present one.
Here, however,
we accepted the simplest parameterization scheme
$p_\parallel=p_{\perp1}=R p_{\perp2}$ involving a single tunable parameter $R$.

\subsection{Correlation-length critical exponent $\nu$}

In this subsection, we study the criticality at $p=p_c$.
In Fig. \ref{figure4}, we plotted the approximate
correlation-length critical exponent \cite{Nightingale76},
\begin{equation}
\nu(L_1,L_2)=
    \left.
 \ln (L_1/L_2) / 
   \ln \left( 
        \frac{\partial (\xi(L_1)/L_1) }{\partial p}/
        \frac{\partial (\xi(L_2)/L_2) }{\partial p} 
       \right) 
     \right|_{p=p_c} ,
\end{equation}
for  $(2/(L_1+L_2))^2$.
Here, we set $p_c=0.0958$ and $R=3.3$.
The symbols $+$, $\times$, and $*$ show that the differences of the system sizes
are
$N_1-N_2=1$, $2$, and $3$, respectively.
Note that this formula contains $p$ derivative,
which is readily calculated with the transfer-matrix method very precisely.
(The transfer-matrix data are free from the statistical error.)

We see that the data align rather satisfactorily.
The least-square fit to these data yields an extrapolated value 
$\nu=0.812(15)$ to the thermodynamic limit $L\to\infty$.
Similarly, we obtain $\nu=0.813(21)$ by omitting the $L=10$ data.
Thereby, we confirm
that the data are almost converged.

In order to check the reliability of $\nu$ further, 
we try to manage alternative extrapolation schemes:
First, we replace the abscissa in Fig. \ref{figure3} 
with the refined one $(2/(L_1+L_2))^{\omega+1/\nu}$
\cite{Binder81}, where we
used $\omega=1.61(5)$ and $\nu=0.89(2)$ reported in Ref. \cite{Lorenz98}.
Thereby, we arrive at $\nu=0.811(15)$. This result indicates the stability of $\nu$ with
respect to $p_c$. 
Second, replacing the abscissa in Fig. \ref{figure4} with $(2/(L_1+L_2))^\omega$,
we obtain $\nu=0.853(19)$.
Actually, this refined extrapolation yields an ``improved'' value for $\nu$.
However, for the sake of self-consistency, we do not accept this refined extrapolation 
method, and consider it as a reference.
Lastly, setting the values of the anisotropy parameter as $R=2.8$ and $R=3.8$,
we obtain $\nu=0.853(27)$ and $\nu=0.771(11)$, respectively.
These results again confirm the stability of $\nu$ satisfactorily.

Recollecting these results, we estimate the correlation-length critical exponent
as,
\begin{equation}
\nu=0.81(5)    ,
\end{equation}
with an expanded error margin.

Let us recollect a number of recent estimates for $\nu$ determined with other approaches:
From the series expansion method, Dunn {\it et al.} \cite{Dunn75} obtained $\nu=0.83(5)$.
On the other hand, the Monte-Carlo studies have reported $\nu=0.89(2)$ \cite{Lorenz98},
$\nu=0.8765(16)$ \cite{Ballesteros99},
$\nu=0.893(40)$ \cite{Tomita02}, and $\nu=0.868(11)$ \cite{Martins03}.
%The followings are the Monte-Carlo studies.
%Lorenz and Ziff obtained $\nu=0.89(2)$ \cite{Lorenz98}.
%Similar conclusion $\nu=0.893(40)$ was reported by Tomita and Okabe \cite{Tomita02}.
%Martins and Plascak obtained a somewhat refined value $\nu=0.868(11)$ \cite{Martins03}.
These estimates and ours are consistent with each other within the error margins.
Nevertheless, we stress that our motivation is not necessarily directed to the accurate estimation
of the critical indices.
In the last section, we address an extended remark on the potential applicability of our scheme
and future perspective.

Lastly, in Fig. \ref{figure5}, we present the approximate $\beta$ function \cite{Roomany80},
\begin{equation}
\label{beta_function}
\beta(L_1,L_2)=\frac{ 1- \ln (\xi(L_1)/\xi(L_2)) / \ln(L_1/L_2) }
    {   \sqrt{ \frac{\partial \xi(L_1)}{\partial p}
               \frac{\partial \xi(L_2)}{\partial p}/\xi(L_1)/\xi(L_2) } }  ,
\end{equation}
for $R=3.3$.
(Note that this formula also contains the derivatives, and
it is hardly accessible by other approaches.)
The symbols $+$, $\times$, and $*$ show that the pairs of 
system sizes are $(N_1,N_2)=(6,8)$, $(7,9)$, and $(8,10)$, respectively.
The beta function provides rich information on the overall feature of the criticality.
The zero point $\beta(p)|_{p=p_c}=0$ indicates the location of the transition point $p_c$,
and the slope at the transition point yields the inverse of $\nu$.
In the figure, we presented a slope $-(p-p_c)/\nu$ with
$p_c=0.0958$ and $\nu=0.81$ determined above.
The slope well describes the behavior of the beta function in the vicinity of $p=p_c$.
However, in a closer look, the beta function bends convexly, indicating that non-negligible 
corrections to scaling do exist.
Possibly, such severe corrections are reflected in Fig. \ref{figure4}, 
where we observe pronounced finite-size corrections to $\nu$.
However, these corrections are fairly systematic so that
we could manage the extrapolation to $L\to\infty$ rather unambiguously.
This is an advantage of Novotny's method, with which
a variety of system sizes are available even for the case of $d=3$.

\subsection{Isotropic case $p_\parallel=p_{\perp 1}=p_{\perp 2}$}

In Fig. \ref{figure6}, we plotted the beta function 
$\beta(L_1,L_2)$ for $R=1$ (isotropic case).
The symbols $+$, $\times$, and $*$ show that the pairs of 
system sizes are $(N_1,N_2)=(6,8)$, $(7,9)$, and $(8,10)$, respectively.
We also presented a slope $-(p-p_c)/\nu$ with $p_c=0.2488126$ and 
$\nu=0.89$ \cite{Lorenz98} as a dashed line.
We see that our data and the slope behave similarly in the vicinity of $p=p_c$.
We obtain the correlation-length critical exponent $\nu \approx 0.64$ from
a pair of $N=8$ and $10$.
However, for small $N$, the data scatter,
and eventually, even the zero point of the beta function disappears.
Because of this irregularity, we cannot
manage systematic extrapolation to the thermodynamic limit.
In this sense, 
the anisotropy parameter $R$ is significant in order to
stabilize the finite-size corrections of the transfer-matrix data
as demonstrated in the preceding subsections.

\section{\label{section4}Summary and discussions}

\begin{figure}
\includegraphics{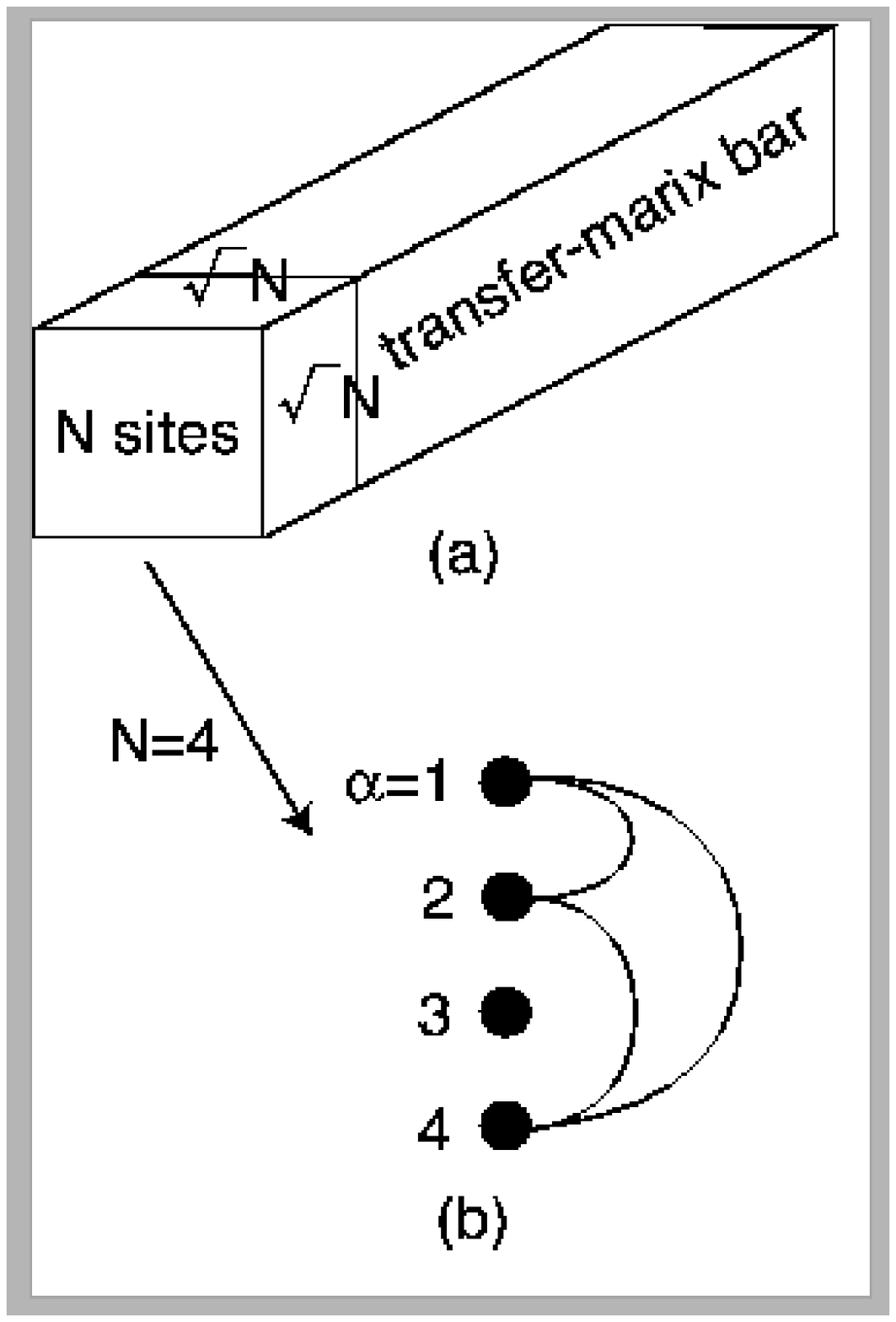}%
\caption{\label{figure1}
(a) A drawing of the transfer-matrix bar.
A unit of the transfer-matrix slice consists of $N$ sites.
(b) An example of connectivity among the $N(=4)$ sites.
Such a connectivity is represented by the matrix notation, Eq. (\ref{matrix_representation}).
}
\end{figure}

\begin{figure}
\includegraphics{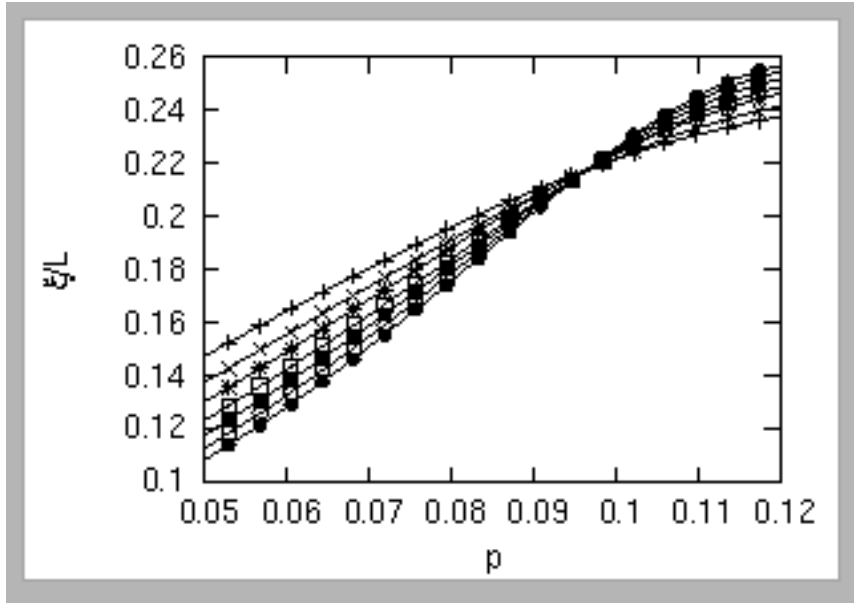}%
\caption{\label{figure2}
The scaled correlation length $\xi/L$ is plotted for
the percolation probability $p$ with the fixed anisotropy parameter $R=3.3$.
The symbols 
$+$, $\times$, $*$, $\Box$, $\blacksquare$,  $\circ$, and $\bullet$
denote the system sizes $N=4$, $5$, $6$, $7$, $8$, $9$, and $10$, respectively;
note that the relation $L=\sqrt{N}$ holds.
From the intersection point of these curves, we read off the
location of the critical point (percolation threshold) as $p_c \approx 0.096$.
}
\end{figure}

\begin{figure}
\includegraphics{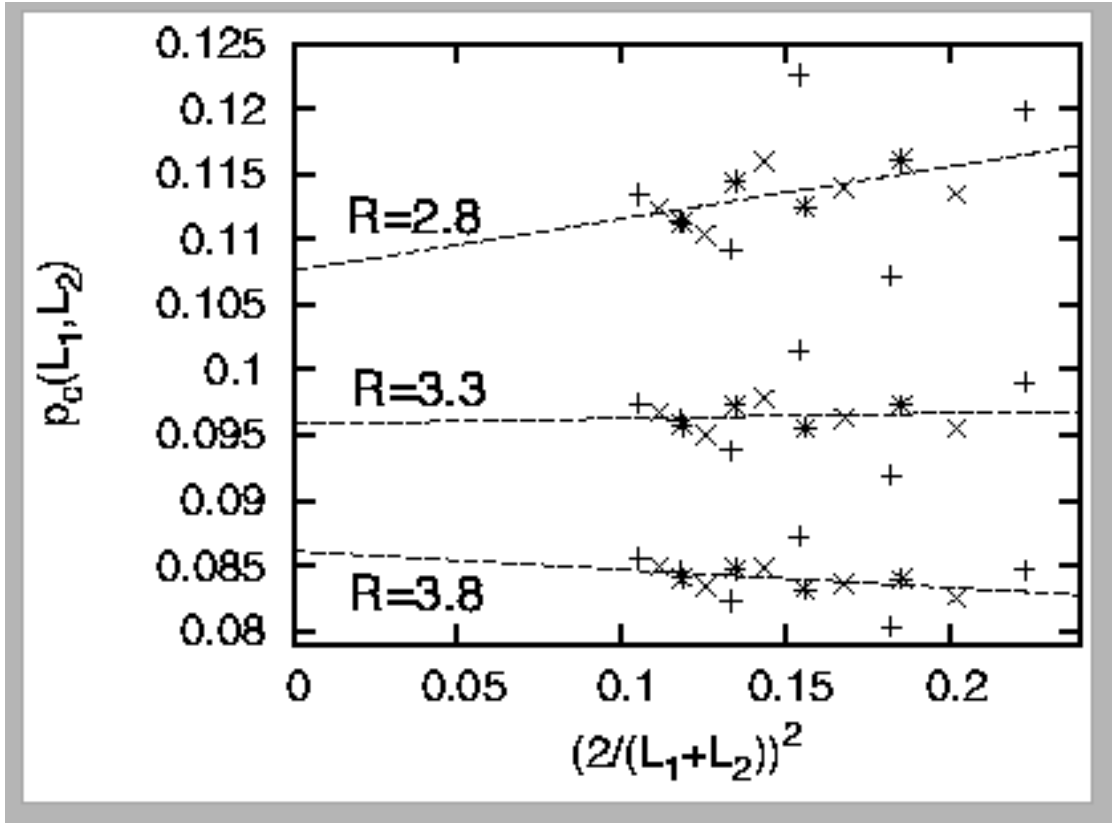}%
\caption{\label{figure3}
The approximate critical point $p_c(L_1,L_2)$ is plotted for 
$(2/(L_1+L_2))^2$ and the various values of the anisotropy parameter $R=2.8$, $3.3$, and $3.8$.
The symbols $+$, $\times$, and $*$ denote the differences of the system sizes
$N_1-N_2=1$, $2$, and $3$, respectively.
We also presented the slopes with the least-square fit to the data as 
the dashed lines.
We see that the finite-size corrections to $p_c$ are suppressed on setting $R=3.3$.
}
\end{figure}

\begin{figure}
\includegraphics{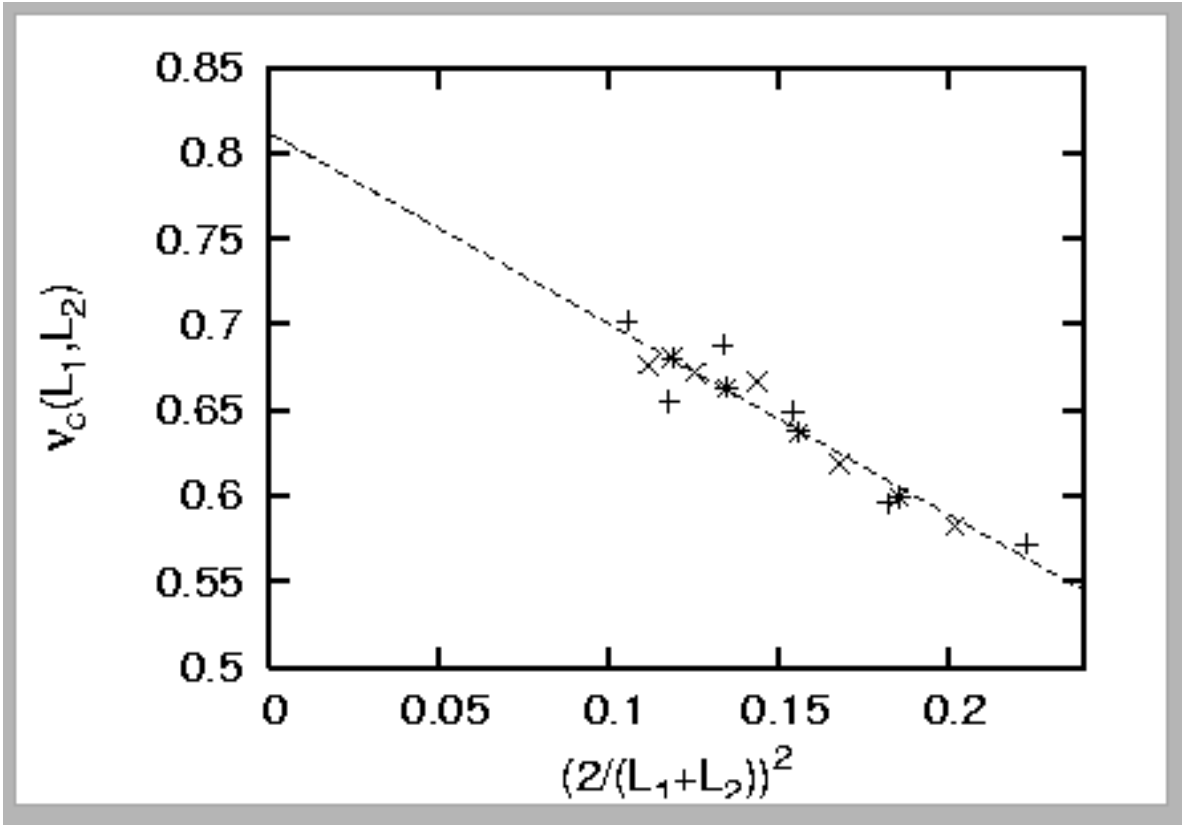}%
\caption{\label{figure4}
The approximate correlation-length critical exponent $\nu(L_1,L_2)$ is
plotted for $(2/(L_1+L_2))^2$ with the fixed anisotropy parameter $R=3.3$.
The symbols $+$, $\times$, and $*$ denote the differences of the system sizes
$N_1-N_2=1$, $2$, and $3$, respectively.
The least-square fit to these data yields $\nu=0.812(15)$ in
the limit $L\to\infty$.
}
\end{figure}

\begin{figure}
\includegraphics{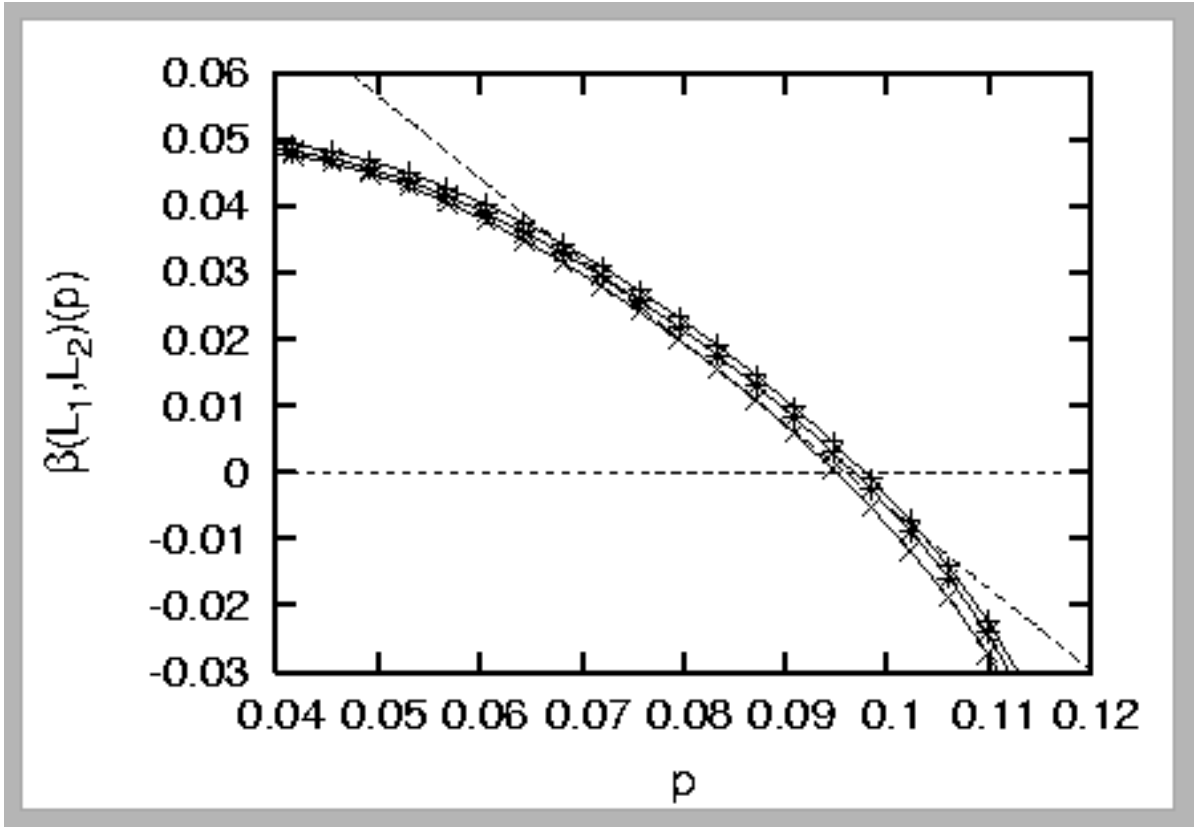}%
\caption{\label{figure5}
The
approximate beta function $\beta(L_1,L_2)$ (\ref{beta_function}) is plotted for 
the percolation probability $p$ with the fixed anisotropy parameter $R=3.3$.
The symbols $+$, $\times$, and $*$ show that the pairs of the system sizes 
are $(N_1,N_2)=(6,8)$, $(7,9)$, and $(8,10)$, respectively. 
We also presented a slope $-(p-p_c)/\nu$ with $p_c=0.0958$ and $\nu=0.812$ determined
in Figs. \ref{figure3} and \ref{figure4}.
}
\end{figure}

\begin{figure}
\includegraphics{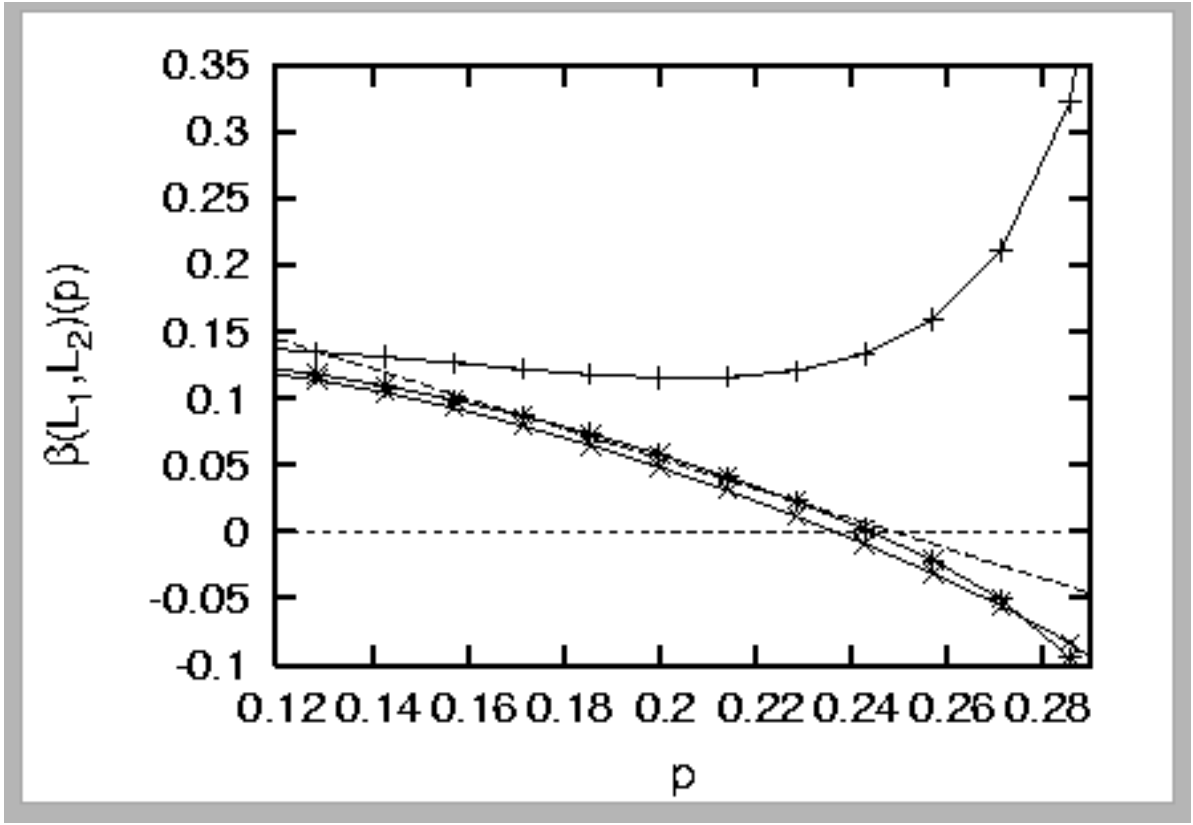}%
\caption{\label{figure6}
The
approximate beta function $\beta(L_1,L_2)$ (\ref{beta_function}) is plotted for 
the percolation probability $p$ at $R=1$ (isotropic point).
The symbols $+$, $\times$, and $*$ show that the pairs of the system sizes 
are $(N_1,N_2)=(6,8)$, $(7,9)$, and $(8,10)$, respectively. 
We also presented a slope $-(p-p_c)/\nu$ with $p_c=0.2488126$ and $\nu=0.89$ reported
in Ref. \cite{Lorenz98} as a dashed line.
}
\end{figure}

We developed a transfer-matrix formalism for the $d=3$ bond percolation.
Our formalism is based on Novotny's idea \cite{Novotny90}, which has been applied 
to the Ising models in high dimensions $d \le 7$.
We demonstrated that his idea is also applicable to the $d=3$ bond percolation.
A key ingredient of his method is that we can treat arbitrary number of
sites constituting a unit of the transfer-matrix slice; see Fig. \ref{figure1}.

Diagonalizing the transfer matrix for the system sizes $N=4,5,\dots,10$,
we studied the criticality of the percolation transition.
We found that the numerical data are well described by the finite-size-scaling theory,
and thereby, we obtained an estimate for the correlation-length critical exponent 
$\nu=0.81(5)$.
Here, we tuned the anisotropy parameter to $R=3.3$ in order to reduce the
finite-size corrections.
Because the system size along the transfer-matrix direction is infinite,
it is by no means necessary to consider the isotropic limit $R=1$ specifically.

%The bond percolation is a limiting case 
%($q \to 1$) of the $q$-state Potts model \cite{Fortuin72}.
%In other words, the present method has a potential applicability
%to the $d=3$ Potts model with continuously variable parameter $q$.
%In fact, in the case of $d=2$,
%the general-$q$-state Potts model has been studied extensively 
%with the transfer-matrix method \cite{Blote82}.
%Actually, the zeros of the partition function in the complex-$q$ plane
%is of current interest \cite{Kim00,Kim01}.
%It would be desirable that similar consideration is given to the $d=3$
%Potts model with ${}^\forall q$ by extending the present transfer-matrix simulation scheme.

The aim of this paper is to demonstrate an applicability of the transfer-matrix method
to the geometrical problem such as the percolation {\em even for} $d=3$.
As mentioned in Introduction, the transfer-matrix method has some advantages over the Monte-Carlo method.
Actually, as for the $d=2$ percolation, 
the transfer-matrix approach \cite{Blote82} has made a unique contribution,
although its accuracy as to the critical indices is not particularly superior to that of Monte Carlo.
Lastly, let us address a remark on the advantage of our approach:
For example, according to Fortuin and Kasteleyn \cite{Fortuin72}, 
the $q$-state-Potts model admits a ``geometrical'' representation
in terms of the bond percolation.
Namely, based on the percolation framework, one is able to
extend the integral number $q$ to a continuously variable one.
In fact, in $d=3$, an extensive Monte-Carlo study \cite{Lee91} reports
an existence of the critical value $q_c=2.45(10)$,
above which the magnetic transition becomes discontinuous; see also Ref. \cite{Gliozzi02}.
However, the nature of this singularity is not fully understood, because the Monte Carlo method cannot
deal with the complex-$q$ number (due to the negative sign problem).
On the contrary, {\em the transfer-matrix method is free this difficulty.}
Actually, with the transfer-matrix method, 
in $d=2$, in the complex-$q$ plane, the distribution of zeros (of the partition function) 
was investigated \cite{Kim00,Kim01,Chang01}.
As for $d=3$, however, similar consideration has not yet been done due to
lack of efficient algorithm.
Our scheme meets such a requirement.
Moreover, the low-lying spectrum of the Potts model in $d=3$ is of fundamental significance \cite{Wang93},
and the problem is also remained unsolved.
In this sense, our scheme provides a step toward a series of such longstanding problems,
to which the Monte-Carlo method does not apply.
An effort toward this direction is in progress, and it will be addressed in future study.

\begin{acknowledgments}
This work is supported by a Grant-in-Aid for
Young Scientists
(No. 15740238) from Monbukaagakusho, Japan.
\end{acknowledgments}


\end{document}